\documentstyle[proceedings]{crckapb} 
\begin{opening}
\title{INTRODUCTORY LECTURE}
\author{MARTIN J. REES}
\institute{Institute of Astronomy\\
           Madingley Road, Cambridge, CB3 0HA}
\end{opening}
\begin{document}

\begin{abstract}
This brief introduction to the NATO ASI offers comments on current
 controversies, on the limits and prospects of cosmology in the coming
 decade, and on how the `sociology' of our subject may change in the
 decades beyond that.
\end{abstract}

\section{Preamble}

    First, a historical perspective:

    What would a conference on `cosmology' have been like in earlier
decades?  In the first half of the century the agenda would have been
almost solely theoretical. The standard models date from the 1920s and
1930s; the Hubble expansion was first claimed in 1929, but not until
the 1950s was there any prospect of discriminating among the various
models, Indeed, there was  even then little quantitative data on how
closely any isotropic homogeneous model fitted the actual universe.

     A cosmology meeting in the 1950s would have focussed on the
question: is there evidence for evolution, or is the universe in a
steady state? Key protagonists on the theoretical side would have
included Hoyle and Bondi.  Ryle would have been arguing that counts of
radio sources -- objects so powerful that many lay beyond the range of
optical telescopes -- already offered evidence for cosmic evolution;
and Sandage would have advocated the potential of the Mount Palomar
200 inch telescope for extending the Hubble diagram far enough to
probe the deceleration.  Intimations from radio counts that the
universe was indeed evolving, were strengthened after 1963 by the
redshift data on quasars.

     The modern era of physical cosmology of course began in 1965,
when the discovery of the microwave background brought the early
`fireball phase' into the realm of empirical science, and the basic
physics of the `hot big bang' was worked out. (The far earlier
contributions by Gamow's associates, Alpher and Herman, continued,
however, to be under-appreciated). There was also substantial
theoretical work on anisotropic models, etc.  Throughout the 1970s
this evidence for the `hot big bang' firmed up, as did the data on
light elements, and their interpretation as primordial relics.

    Theoretical advances in the 1980s gave momentum to the study of
the ultra-early universe, and fostered the `particle physics
connection': the sociological mix of cosmologists changed. There was
intense discussion of inflationary models, non-baryonic matter, and so
forth.

    Here we are in the late 1990s with a still larger and more diverse
community. The pace of cosmology has never been faster. We're
witnessing a crescendo of discoveries that promises to continue into
the next millennium.  This is because of a confluence of developments:
 
   1. $\underline{\hbox{The microwave background fluctuations}}$ : these are now being
probed with enough sensitivity to provide crucial tests of inflation
and discrimination among different models.

   2.  $\underline{\hbox{The high-redshift universe}}$: the Hubble Space Telescope (HST)
has fulfilled §its potential; two Keck Telescopes have come on line,
along with the first VLT telescope, Subaru, and the first Gemini
telescope. These have opened up the study of `ordinary' galaxies right
back to large redshifts, and to epochs when they were newly formed. In
the coming year, three new X-ray telescopes will offer higher
resolution and higher sensitivity for the study of distant galaxies
and clusters.

      3.  $\underline{\hbox{Large scale clustering and dynamics}}$: big surveys currently
in progress are leading to far larger and more reliable samples, from
which we will be able to infer the quantitative details of how
galaxies of different types are clustered, and how their motions
deviate from Hubble flow.  Simultaneously with this progress, there
have been dramatic advances in computer simulations.  These now
incorporate realistic gas dynamics as well as gravity.

  4.  $\underline{\hbox{Developments in fundamental physics}}$ offer important new
speculative insights, which will certainly figure prominently in our
discussions of the ultra-early universe.

  It is something of a coincidence  -- of technology and funding --  that the
impetus on all these fronts has been more or less concurrent.

      Max Planck claimed that theories are never abandoned until their
proponents are all dead: that's too cynical, even in cosmology!  Some
debates  have been settled; some earlier issues are no
longer controversial; some of us change our minds (quite frequently, 
sometimes).   And as the consensus advances, new questions
which couldn't even have been posed in earlier decades are now being
debated.  This conference's agenda is therefore a `snapshot' of
evolving knowledge, opinion and speculation.
 
      Consider the following set of assertions -- a typical utterance
of the r-m-s cosmologist whom you might encounter on a Cambridge
street.  \hfill\break
Our universe is expanding\hfill\break
from a hot big bang\hfill\break
in which the light elements were synthesised.\hfill\break
There was a  period of inflation,\hfill\break
which led to a `flat' universe today.\hfill\break
Structure was `seeded' by gaussian irregularities,\hfill\break
which are the relics of quantum fluctuations,\hfill\break
and  the large-scale dynamics is dominated by `cold' dark matter.\hfill\break
but   $\Lambda$  (or quintessence) is dynamically dominant.

     I've written it like nine lines of `free verse' to highlight how
some claims are now quite firm, but others are fragile and
tentative. Line one is now a quite ancient belief: it would have rated
99 percent confidence for several decades. Line two represents more
recent history, but would now be believed almost equally strongly.
So, probably, would line three, owing to the improved observations of
abundances, together with refinements in the theory of cosmic
nucleosynthesis.  The concept of inflation is now 20 years old; most
cosmologists suspect that it is was indeed a crucial formative process
in the ultra-early universe, and this conference testifies to the
intense and sophisticated theorising that it  still stimulates.

   Lower down in my list of statements, the confidence level drops
below 50 percent. The `stock' in some items -- for instance CDM models
, which have had several `deaths' and as many `resurrections' -- is
volatile, fluctuating year by year! The most spectacular 'growth
stock' now (but for how long?) is the cosmological constant, lambda.

\section{The Cosmological Numbers}

     Traditionally, cosmology was the quest for a few numbers. The
first were H, q, and $\Lambda$.  Since 1965 we've had another : the
baryon/photon ratio. This is believed to result from a small
favouritism for matter over antimatter in the early universe --
something that was addressed in the context of `grand unified
theories' in the 1970s. (Indeed, baryon non-conservation seems a
prerequisite for any plausible inflationary model.  Our entire
observable universe, containing at least $10^{79}$ baryons, could not have
inflated from something microscopic if baryon number were strictly
conserved)
  
  In the 1980s non-baryonic matter became almost a natural
expectation, and $\Omega_b /\Omega_{\rm CDM}$ is another fundamental number .

    Another specially important dimensionless number, ${\rm Q}$,  tells us how
smooth the universe is. It's measured by

--- The Sachs-Wolfe fluctuations in the microwave background 

--- the gravitational binding energy of clusters as a fraction of
their rest mass

-- or by the square of the typical scale of mass- clustering as a
fraction of the Hubble scale.
 
    It's of course oversimplified to represent this by a single number
Q, but insofar as one can, its value is pinned down to be
$10^{-5}$. (Detailed discussions introduce further numbers: the ratio of
scalar and tensor amplitudes, and quantities such as the `tilt', which
measure the deviation from a pure scale-independent Harrison-Zeldovich
spectrum.)

     What's crucial is that Q is small.  Numbers like $\Omega$ and H
are only well-defined insofar as the universe possesses `broad brush'
homogeneity -- so that our observational horizon encompasses many
independent patches each big enough to be a fair sample. This wouldn't
be so, and the simple Friedmann models wouldn't be useful
approximations, if Q weren't much less than unity.  Q's smallness is
necessary if the universe is to look homogeneous.  But it isn't,
strictly speaking, a sufficient condition -- a luminous tracer that
didn't weigh much could be correlated on much larger scales without
perturbing the metric.  Simple fractal models for the luminous matter
are nonetheless, as Lahav will discuss, strongly constrained by other
observations such as the isotropy of the X-ray background, and of the
radio sources detected in deep surveys.

\section{How confident can we be of our models?}

    If our universe has indeed expanded, Friedmann-style, from an
exceedingly high density, then after the first $10^{-12}$ seconds
energies are within the range of accelerators. After the first
millisecond -- after the quark-hadron transition -- conditions are so
firmly within the realm of laboratory tests that there are no crucial
uncertainties in the microphysics (though we should maybe leave our
minds at least ajar to the possibility that the constants may still be
time-dependent). And everything's still fairly uniform --
perturbations are still in the linear regime.

   It's easy to make quantitative predictions that pertain to this
intermediate era, stretching from a millisecond to a million
years. And we've now got high-quality data to confront them with.  The
marvellous COBE `black body' pins down the microwave background
spectrum to a part in 10,000. The `hot big bang' has lived dangerously for thirty years: it could have been shot down by (for instance) the discovery of a nebula with zero helium, or of a stable neutrino with keV mass; but nothing like this has happened.  The debate
(concurrence or crisis?) now focuses on 1 per cent effects in helium
spectroscopy, and on traces of deuterium at very high redshifts. The
case for extrapolating back to a millisecond is now compelling and
battle-tested. Insofar as there's a `standard model' in cosmology,
this is now surely part of it.

      When the primordial plasma recombined, after half a million
years, the black body radiation shifted into the infrared, and the
universe entered, literally, a dark age. This lasted until the first
stars lit it up again. The basic microphysics remains, of course,
straightforward. But once non-linearities develop and bound systems
form, gravity, gas dynamics, and the physics in every volume of Landau
and Lifshitz, combine to unfold the complexities we see around us and
are part of.

    Gravity is crucial in two ways.  It first amplifies `linear'
density contrasts in an expanding universe; it then provides a
negative specific heat so that dissipative bound systems heat up
further as they radiate. There's no thermodynamic paradox in evolving
from an almost structureless fireball to the present cosmos, with huge
temperature differences between the 3 degrees of the night sky, and
the blazing surfaces of stars.

   It is feasible to calculate all the key cosmic processes that
occurred between (say) a millisecond and a few million years: the
basic physics is `standard' and (according at least to the favoured
models) everything is linear.  The later universe, after the dark age
is over, is difficult for the same reason that all environmental
sciences are difficult.

   The whole evolving tapestry is, however, the outcome of initial
conditions (and fundamental numbers) imprinted in the first
microsecond -- during the era of inflation and baryogenesis, and
perhaps even on the Planck scale . This is the intellectual habitat of
specialists in quantum gravity, superstrings, unified theories, and
the rest.

      So cosmology is a sort of hybrid science. It's a `fundamental'
science, just as particle physics is. But it's also the grandest of
the environmental sciences.  This distinction is useful, because it
signals to us what levels of explanation we can reasonably expect. The
first million years is described by a few parameters: these numbers
(plus of course the basic physical laws) determine all that comes
later. It's a realistic goal to pin down their values.  But the cosmic
environment of galaxies and clusters is now messy and complex -- the
observational data are very rich, but we can aspire only to an
approximate, statistical, or even qualitative `scenario', rather like
in geology and paleontology.

  The relativist Werner Israel likened this dichotomy to the contrast
between chess and mudwrestling. The participants in this meeting would
seem to him, perhaps. an ill-assorted mix of extreme refinement and
extreme brutishness (just in intellectual style, of course!).

\section{Complexities of Structure, and Dark Matter}

\subsection{Prehistory of Ideas on Structure Formation}

Since we are meeting in the Isaac Newton Institute, it's fitting to recall that Newton himself envisaged structures forming via `gravitational instability'. In an often-quoted letter  to Richard Bentley, the Master of Trinity College, he wrote:

``If all the matter of the universe were evenly scattered throughout all the heavens, and every particle had an innate gravity towards all the rest, and ... if the matter were evenly dispersed throughout an infinite space, it could never convene into one mass, but some of it would convene into one mass and some into another, so as to make an infinite number of great masses, scattered at great distances from one another throughout all that infinite space. And thus might the sun and fixed stars be formed. ... supposing the matter to be of a lucent nature.'' 

(It would of course be wishful thinking to interpret his last remark as a premonition of dark matter!)

\subsection{The role of simulations}

 Our view of cosmic evolution is, like Darwinism, a compelling general
scheme.  As with Darwinism, how the whole process got started is still
a mystery. But cosmology is simpler because, once the starting point
is given, the gross features are predictable. The whole course of
evolution isn't, as in biology, sensitive to `accidents'.  All large
patches that start off the same way, end up statistically similar.

  That's why simulations of structure formation are so
important. These have achieved higher resolution, and incorporate gas
dynamics and radiative effects as well as gravity.  They show how
density contrasts grow from small-amplitude beginnings; these lead,
eventually, to bound gas clouds and to internal dissipation.
 
    Things are then more problematical. We're baffled by the details
of star formation now, even in the Orion Nebula. What chance is there,
then, of understanding the first generation of stars, and the
associated feedback effects?  In CDM-type models, the very first stars
form at redshifts of 10-20 when the background radiation provides a
heat bath of up to 50 degrees, and there are no heavy elements. There
may be no magnetic fields, and this also may affect the initial mass
function. We also need to know the efficiency of star formation, and
how it depends on the depth of the potential wells in the first
structures.

    Because these problems are too daunting to simulate {\it ab
initio}, we depend on parameter-fitting guided by observations. And
the spectacular recent progress from 10-metre class ground based
telescopes and the HST has been extraordinarily important here.

\subsection{Observing high redshifts}

   We're used to quasars at very high redshifts. But quasars are rare
and atypical -- we'd really like to know the history of matter in
general.  One of the most important advances in recent years has been
the detection of many hundreds of galaxies at redshifts up to (and
even beyond) 5.  Absorption due to the hundreds of clouds along the
line of sight to quasars probes the history of cosmic gas in exquisite
detail, just as a core in the Greenland ice-sheet probes the history
of Earth's climate.
 
  Quasar activity
reaches a peak at around $z= 2.5$. The rate of star formation may peak
at somewhat smaller redshifts (even though the very first starlight
appeared much earlier) But for at least the last half of its history,
our universe has been getting dimmer.  Gas gets incorporated in
galaxies and `used up' in stars -- galaxies mature, black holes in
their centres undergo fewer mergers and are starved of fuel, so AGN
activity diminishes.
 
    That, at least, is the scenario that most cosmologists accept. To
fill in the details will need better simulations.  But, even more, it
will need better observations. I don't think there is much hope of
`predicting' or modelling the huge dynamic range and intricate
feedback processes involved in star formation.  A decade from now,
when the Next Generation Space Telescope (NGST) flies, we may know the
main cosmological parameters, and have exact simulations of how the
dark matter clusters. But reliable knowledge of how stars form, when
the intergalactic gas is reheated, and how bright the first
`pregalaxies' are will still depend on observations. The aim is get a
consistent model that matches not only  all we know about galaxies at the
present cosmic epoch, but also the increasingly detailed snapshots of what
they looked like, and how they were clustered, at all earlier times.

   But don't be too gloomy about the messiness of the `recent'
universe.  There are some `cleaner' tests.  Simulations can reliably
predict the present clustering and large-scale distribution of
non-dissipative dark matter. This can be observationally probed by
weak lensing, large scale streaming, and so forth, and checked for
consistency with the CMB fluctuations, which probe the linear
precursors of these structures.

\subsection{Dark matter: what, and how much?}    

The nature of the dark matter -- how much there is and what it is --
still eludes us. It's embarrassing that 90 percent of the universe
remains unaccounted for.

This key question may yield to a three-pronged attack:

1. {$\underline{\hbox{\rm Direct detection}}$}. Astronomical searches are underway for `machos'
in the Galactic Halo; and several groups are developing cryogenic
detectors for supersymmetric particles and axions.
 
2. {$\underline{\hbox{\rm Progress in particle physics}}$}. Important recent measurements suggest
that neutrinos have non-zero masses; this result has crucially
important implications for physics beyond the standard model; however
the inferred masses seem too low to be cosmologically important.  If
theorists could pin down the properties of supersymmetric particles,
the number of particles that survive from the big bang could be calculated just as
we now calculate the helium and deuterium made in the first three
minutes.  Optimists may hope for progress on still more exotic
options.

3. {$\underline{\hbox{\rm Simulations of galaxy formation and large-scale structure}}$}.  When
and how galaxies form, the way they are clustered, and the density
profiles within individual systems, depend on what their
gravitationally-dominant constituent is, and are now severely
constraining the options.

\section{Steps  Beyond the  Simplest Universe: Open Models, ${\bf \Lambda}$, etc.} 

\subsection{The case for $\Omega < 1$}

   Everyone agrees that the `simplest' universe would be a flat
Einstein-de Sitter model. But we shall hear several claims during the
present meeting that this model is now hard to reconcile with the
data.  Several lines of evidence suggest that gravitating CDM
contributes substantially less than $\Omega_{\rm CDM} = 1$.  The main
lines of evidence are

(i) The baryon fraction in clusters is 0.15-0.2, On the other hand,
the baryon contribution to omega is now pinned down by deuterium
measurements to be around $\Omega_b = 0.015 h^2$, where $h$ is the Hubble 
constant in units of 100 km/sec/Mpc. If clusters are a fair
sample of the universe, then this is incompatible with a dark matter
density high enough to make $\Omega =1$.

(ii) The presence of clusters of galaxies with $z=1$ is hard to
reconcile with the rapid recent growth of structure that would be
expected if $\Omega_{\rm CDM}$ were unity.

(iii) The Supernova Hubble diagram (even though the case for actual
acceleration may not be compelling) seems hard to reconcile with the
large deceleration implied by an Einstein-de Sitter model.

(iv) The inferred ages of the oldest stars are only barely consistent
with an Einstein-de Sitter model, for the favoured choices of Hubble
constant.

\subsection{Open universe, or vacuum energy?}

  The two currently-favoured options seem to be:

    (A) an open model, or else

 (B) a flat model where vacuum energy (or some negative-pressure
component that didn't participate in clustering) makes up the balance.

If the universe is a more complicated place than some people hoped,
which of these options is the more palatable?  Opinions here may
differ: How `contrived' are the open-inflation models? Is it even more
contrived that the vacuum-energy should have the specific small value
that leads it to start dominating just at the present epoch?

   Either of these models involves a specific large number. In case
(A) this is the ratio of the Robertson Walker curvature scale to the
Planck scale; in (B) it is the ratio of vacuum energy to some other
(much higher) energy density.  At present, (A) seems to accord less
well than (B) with the data. In particular, the angular scale of the
`doppler peaks' in the CMB angular fluctuations seems to favour a flat
universe; and the supernova Hubble diagram indicates an actual
acceleration, rather than merely a slight deceleration (as would be
expected in the open model).

    We will certainly hear a great deal about the mounting evidence
for $\Lambda$ (or one of its time-dependent generalisations): the claimed
best fit to all current data suggests a non-zero energy in the
vacuum. However we should be mindful of  the current large scatter in all CMB
measurements relevant to the doppler peak, and the various
uncertainties (especially those that depend on composition, etc.) in
the supernovae from which a cosmic acceleration has been inferred. I
think the jury is still out. However, CMB experiments are developing
fast, and the high-$z$ supernova sample is expanding fast too; so within
two years we should know whether there is a vacuum energy, or whether
systematic intrinsic differences between high-$z$ and low-$z$ supernovae
are large enough to render the claims spurious.  (On the same
timescale we should learn whether the Universe actually is flat).

\subsection{The History of $\Lambda$}

  I wouldn't venture bets on the final status of $\Lambda$. It is
nonetheless interesting to recall its history. $\Lambda$ was of course
introduced by Einstein in 1917 to permit a static unbounded
universe. After 1929, the cosmic expansion rendered Einstein's
motivation irrelevant. However, by that time de Sitter had already
proposed his expanding $\Lambda$-dominated model. In the 1930s, Eddington
and Lemaitre proposed that the universe had expanded (under the action
of the $\Lambda$-induced repulsion) from an initial Einstein
state. $\Lambda$ fell from favour after the 1930s: relativists disliked
it as a `field' acting on everything but acted on by nothing.  A brief
resurgence in the late 1960s was triggered by a (now discredited)
claim for a pile-up in the redshifts of quasars at a value of $z$
slightly below 2. (The CMB had already convinced most people that  the
universe emerged from a dense state, rather than from an Einstein
static model, but it could have gone through a coasting or loitering
phase where the expansion almost halted. A large range of affine
distance would then correspond to a small range of redshifts, thereby
accounting for a `pile up' at a particular redshift. It was also
noted that this model offered more opportunity for small-amplitude
perturbations to grow.)

  The `modern' interest in $\Lambda$ stems from its interpretation as an
vacuum energy. This leads to the reverse problem: Why is $\Lambda$ $\sim 120$ powers of 10 smaller than its `natural' value, even though
the effective vacuum density must have been very high in order to
drive inflation?
   The interest has of course been hugely boosted recently, through
the claims that the Hubble diagram for Type 1A supernovae indicates an
acceleration.  

(If $\Lambda$ is fully resurrected, it will be a great
`coup' for de Sitter. His model, dating from the 1920s, not only
describes the dynamics throughout the huge number of `e-foldings' during inflation, but also describes future  aeons of our cosmos
with increasing accuracy. Only for the 50--odd decades of logarithmic
time between the end of inflation and the present does it need
modification!).
  
\section{Inflation and the Very Early Universe}

  Numbers like $\Omega, \left(\Omega_b/\Omega_{\rm CDM}\right), \, \Lambda$ and Q
are determined by physics as surely as the He and D abundances -- it's
just that the conditions at the ultra-early eras when these numbers
were fixed are far beyond anything we can experiment on, so the
relevant physics is itself still conjectural.

 The inflation concept is the most important single idea. It suggests
why the universe is so large and uniform -- indeed, it suggests why it
is expanding. It was compellingly attractive when first proposed, and
most cosmologists (with a few eminent exceptions like Roger Penrose)
would bet that it is, in some form, part of the grand cosmic scheme.
The details are still unsettled.  Indeed, cynics may feel that, since
the early 1980s, there've been so many transmogrifications of
inflation -- old, new, chaotic, eternal, and open -- that its
predictive power is much eroded. (But here again extreme cynicism is
unfair.)

    We'll be hearing some discussion of whether inflationary models
can `naturally' account for the fluctuation amplitude ${\rm Q} = 10^{-5}$ ; and,
more controversially, whether it's plausible to have a non-flat
universe, or a present-day vacuum energy in the permissible
range. It's important to be clear about the methodology and scientific
status of such discussion. I comment with great diffidence, because
I'm not an expert here.

      This strand of cosmology may still have unsure foundations, but
it isn't just metaphysics: one can test particular variants of
inflation. For instance, definite assumptions about the physics of the
inflationary era have calculable consequences for the fluctuations
--whether they're gaussian, the ratio of scalar and tensor modes, and
so forth -- which can be probed by observing large scale structure
and, even better, by microwave background observations.  Cosmologists
observe, stretched across the sky, giant proto-structures that are the
outcome of quantum fluctuations imprinted when the temperature was
$10^{15}$ GeV or above. Measurements with the MAP and Planck/Surveyor
spacecraft will surely tell us things about `grand unified' physics
that can't be directly inferred from ordinary-energy experiments.

\section{The  Agenda  10 Years From  Now: a Bifurcated Community?}

\subsection{The next five years}

   The current pace of advance is such that within five years we'll surely 
have made substantial further progress. We will not only agree that the value
of H is known to 10 percent -- we'll agree what that value is.

   We'll know the key parameters (from high-$z$ supernovae, from the
CMB, from high-$z$ observations, and from improved statistics on large
scale clustering and streaming.
    I'd even bet (though maybe I'm being a bit rash here) that we'll
know what the dominant dark matter is.

\subsection{Ten years ahead?}

    If we were to reconvene 10 years from now, what would be the `hot
topics' on the agenda?  The key numbers specifying our universe -- its
geometry, fluctuations and content -- may by then have been pinned
down.  I've heard
people claim that cosmology will thereafter be less interesting-- that
the most important issues will be settled, leaving only the secondary
drudgery of clearing up some details.  I'd like to spend a moment
trying to counter that view.

  It may turn out, of course, that the new data don't fit at all into
the parameter-space that these numbers are derived from.
 (I was tempted to describe this view as `pessimistic'  but of course some people may prefer to
live in a more complicated and challenging universe!).
 But maybe everything will fit the framework, and we will pin down the
contributions to ½ from baryons, CDM, and the vacuum, along with the
amplitude and tilt of the fluctuations, and so forth.  If that
happens, it will signal a great triumph for cosmology -- we will know
the `measure of our universe' just as, over the last few centuries,
we've learnt the size and shape of our Earth and Sun.

   Our focus will then be redirected towards new challenges, as great
as the earlier ones. But the character and `sociology' of our subject
will change: it will bifurcate into two sub-disciplines.  This bifurcation would be analogous to what
actually happened in the field of general relativity 20-30 years
ago. The `heroic age' of general relativity -- leading to the rigorous
understanding of gravitational waves, black holes, and singularities
-- occurred the 1960s and early 1970s. Thereafter, the number of active
researchers in `classical' relativity declined (except maybe in
computational aspects of the subject): most of the leading researchers
shifted either towards astrophysically-motivated problems, or towards
quantum gravity and `fundamental' physics.

    What will be the foci of the two divergent  branches of `post classical'
cosmology we'll be pursuing a decade from now? One will be
`environmental cosmology' -- understanding the evolution of structure,
stars and galaxies. The other will focus on the fundamental physics of
the ultra-early universe (pre-inflation, m-branes, multiverses,
etc). A few words about each of these:

\subsection{Environmental cosmology: long range prospects}

    One continuing challenge will be to explore the emergence of
structure. This is a tractable problem until the first star (or other
collapsed system) forms.  But the huge dynamic range and uncertain
feedback thereafter renders the phenomena too complex for any
feasible simulation.

  To illustrate the uncertainty, consider a basic question such as
when the intergalactic gas was first photoionized.

    There have been many detailed models, but essentially this
requires one photon for each baryon (somewhat more, in fact, to
compensate for recombinations). A hot (O or B) star produces, over its
lifetime, $10^4- 10^5$ photons for each of its constituent baryons; if a
black hole forms via efficient accretion of baryons, the corresponding
number is several times $10^6$.  Thus, only a small amount of material
need collapse into such objects in order to provide enough to ionize
all the remaining baryons. But the key questions, of course, are how
efficiently O-B stars or black holes can form. This depends on the
so-called `initial mass function' (IMF), which determines how much
mass goes into high mass stars (or black holes) compared with the
amount going concurrently into lower-mass stars?  The challenge of
calculating the IMF -- involving gas-dynamical and radiative transfer
calculations over an enormous dynamic range -- may not have been met
even ten years from now. But even if we assume that it has the same
form as now, there is the issue of feedback: do the first stars provide
a heat input (via radiation, stellar winds and supernovae) that
inhibits later ones from forming? More specifically, we can imagine
two options; either (a) all the gas that falls into gravitationally
bound clumps of CDM turns into stars; or (b) one percent turns into
stars, whose winds and supernovae provide enough momentum and energy
to expel the other 99 percent.  In the first case, the 3-sigma peaks
would suffice; in case (b) more typical (1.5 sigma) peaks would be
needed, or else larger and deeper potential wells more able to retain
the gas.

    Even if the clustering of the CDM under gravity could be exactly
modelled, along with the gas dynamics, then as soon as the first stars
form we face major uncertainties that will still be a challenge to the
petaflop simulations being carried out a decade from now.

\subsection{Probing  the Planck era and `beyond'}

    The second challenge would be to firm up the physics of the
ultra-early universe. Perhaps the most `modest' expectation would be a
better understanding of the candidate dark matter particles: if the
masses and cross-sections of supersymmetric particles were known, it
should be possible to predict how many survive, and their
contribution to $\Omega$, with the same confidence as that with which we can
compute primordial nucleosynthesis. Associated with such progress, we
might expect a better understanding of how the baryon-antibaryon
asymmetry arose, and the consequence for $\Omega_b$.
 
   A somewhat more ambitious goal would be to pin down the physics of
inflation. Knowing parameters like Q, the tilt, and the scalar/tensor
ratio will narrow down the range of options.  The hope must be to make
this physics as well established as the physics that prevails after
the first millisecond.

  One question that interests me specially is whether there are
multiple big bangs, and which features of our actual universe are
contingent rather than necessary.  Could the others have different
values of Q,  or different Robertson-Walker  curvature?  Furthermore, will the `final theory'
determine uniquely what we call the fundamental constants of physics
-- particle masses and coupling constants?  Are these `constants'
uniquely specified by some equation that we can eventually write down?
Or are they in some sense accidental features of a phase transition as
our universe cooled -- secondary manifestations of some still deeper
laws governing a whole ensemble of universes?
 
  This might seem arcane stuff, disjoint from `traditional' cosmology
-- or even from serious science.  But my prejudice is to be openminded
about ensembles of universe and suchlike. This makes a real difference
to how I weigh the evidence and place my bets on rival models.
 
      Rocky Kolb's highly readable history `Blind Watchers of the Sky'
reminds us of some fascinating debates that occurred 400 years ago.
Kepler was upset to find that planetary orbits were elliptical.
Circles were more beautiful -- and simpler, with one parameter not
two. But Newton later explained all orbits in terms of a universal law
with just one parameter, G. Had Kepler still been alive then, he'd
surely have been joyfully reconciled to ellipses.

  The parallel's obvious. The Einstein-de Sitter model seems to have
fewer free parameters than any other. Models with low $\Omega$,
non-zero $\Lambda$, two kinds of dark matter, and the rest may seem
ugly. But maybe this is our limited vision. Just as Earth follows an
orbit that is no more special than it needed to be to make it
habitable, so we may realise that our universe is just one of the 
anthropically-allowed members of a grander ensemble.  So maybe we
should go easy with Occam's razor and be wary of arguments that
$\Omega=1$ and $\Lambda=0$ are {\it a priori} more natural and less ad hoc.

   There's fortunately no time to sink further into these murky
waters, so I'll briefly conclude.
          
   A recent cosmology book (not written by anyone at this conference)
was praised, in the publisher's blurb, for `its thorough coverage of
the inflammatory universe'. That was a misprint, of course. But maybe
enough sparks will fly here in the next few days to make it seem a not
inapt description.

 The organisers have chosen a set of fascinating open questions. I
suspect they'll still seem open at the end of this meeting, but we'll
look forward to learning the balance of current opinion, and what bets
people are prepared to place on the various options.
\end{document}